\documentclass[aps,twocolumn,showpacs,preprintnumbers,nofootinbib,prd,superscriptaddress,groupedaddress,10pt]{revtex4-1}

\usepackage{graphicx,amssymb,amsmath,amsthm,amsfonts,epsfig,epsf}
\usepackage[usenames,dvipsnames,svgnames,table]{xcolor}
\usepackage{epstopdf}
\definecolor{darkred}{rgb}{0.5,0,0}
\usepackage{aas_macros}
\usepackage{bm}
\usepackage{dcolumn}
\usepackage[utf8]{inputenc}
\usepackage{latexsym}
\usepackage{rotating}
\usepackage{longtable}

\usepackage{enumerate}
\usepackage{tensor,multirow}
\usepackage{url}
\usepackage[linktocpage]{hyperref}

\begin{document}
\title{Electromagnetism and hidden vector fields in modified gravity theories: \\Spontaneous and induced vectorization}

\author{
Lorenzo Annulli$^{1}$,
Vitor Cardoso$^{1,2}$,
Leonardo Gualtieri$^{3}$
}

\affiliation{${^1}$ Centro de Astrof\'{\i}sica e Gravita\c c\~ao - CENTRA, Departamento de F\'{\i}sica, Instituto
  Superior T\'ecnico - IST, Universidade de Lisboa - UL, Avenida Rovisco Pais 1, 1049-001 Lisboa, Portugal}
\affiliation{${^2}$ Theoretical Physics Department, CERN 1 Esplanade des Particules, Geneva 23, CH-1211, Switzerland}
\affiliation{$^3$ Dipartimento di Fisica, "Sapienza" Universit\`a di Roma and Sezione INFN Roma1, Piazzale Aldo Moro 5,
  00185 Roma, Italy}

\begin{abstract}
In general relativity, Maxwell's equations are embedded in curved spacetime through the minimal prescription, but this could change if strong-gravity modifications are present. We show that with a nonminimal coupling between
gravity and a massless vector field, nonperturbative effects can arise in compact stars. We find solutions describing
stars with nontrivial vector field configurations, some of which are associated with an instability, while others are
not.  The vector field can be interpreted either as the electromagnetic field, or as a hidden vector field weakly
coupled with the standard model.
\end{abstract}

\maketitle
\section{Introduction}
Astronomical measurements on binary pulsar systems~\cite{Lange:2001rn,Antoniadis:2013pzd}, together with observations of
gravitational waves (GWs) emitted by binaries containing neutron stars (NSs), such as GW170817 detected by the LIGO-Virgo
observatory~\cite{TheLIGOScientific:2017qsa}, have improved our knowledge about compact stars. The near future promises
to bring in a wealth of data that will allow us to understand compact objects with an unprecedented precision. Theoretical
predictions about the NS spacetime and the equation of state of matter at such high densities will be compared with
observational data, cementing our knowledge of extreme spacetimes.

Accurate observations concerning regions of spacetime where gravity is ``strong'' will also test general relativity (GR)
and long-held beliefs about how matter behaves in curved spacetime. One example--which will be the focus of this work--is the coupling of vector fields to curved spacetime. In Einstein-Maxwell theory,
the massless vector field $X_\mu$ is embedded in curved spacetime through the standard ``colon-goes-to-semicolon''
rule~\cite{1975pbrg.book.....L}, but there are endless other possibilities. Ultimately, it is up to observations to
determine the appropriate description.  We will focus on a simple and elegant extension proposed by
Hellings-Nordtvedt~\cite{Hellings:1973zz} (HN), detailed below in Eq.~\eqref{eq:action}, in which a further coupling between
the curvature and the vector field is included.

The vector field discussed in this article can be interpreted in different ways, either as (i) the well-known
electromagnetic field, or as (ii) a still unknown vector field, which is ``hidden'' since it is weakly coupled with the
standard model.

Within the interpretation (i), we are studying strong-gravity modifications of the coupling between gravity and the
electromagnetic field. We remark that the effects we are seeking only show up in the presence of a very large spacetime
curvature, such as those in the core of neutron stars, or near the horizon of black holes. Therefore, despite the
enormous accuracy of existing experimental data on the electromagnetic field, the effects studied in this article are
not ruled out by current observations.  In particular, we mainly study the effects of the inclusion of a coupling
$\sim RX_\mu X^\mu$ (where $R$ is the spacetime curvature) in the action (see Eq.~\eqref{eq:action} below).
This coupling resembles a mass term (but with a nonconstant and nonuniform ``mass''). We note that even the existence of a
photon mass has not been definitely ruled out \cite{Eidelman:2004wy,Pani:2012vp}; a
photon-curvature coupling is more elusive, since it shows up only in strong curvature regions.

Within the interpretation (ii), one tries
to enlarge the standard model with as many fields as possible, and question which of those fields can be constrained
with experiments. In this context, the theory \eqref{eq:action} arises naturally, in the sense that (hidden, with
small couplings to the standard model) vectors are a generic prediction of string theory~\cite{Polchinski:1998rq}, and
are promising dark matter candidates~\cite{Essig:2013lka}~\footnote{Generalized theories with vector fields, avoiding
  ghosts and other pathologies, have recently been studied in Ref.~\cite{Heisenberg:2014rta}.}. A natural approach in this
framework is to look for smoking-gun effects of such new fields and couplings. Note that within this interpretation, the
vector field has a double role, as a ``matter'' field belonging to a hidden sector of the standard model, and
as a carrier, together with the spacetime metric, of the gravitational interaction; thus, HN gravity can be considered
as an example of a ``vector-tensor'' theory.

In the context of scalar-tensor theories, a nonminimal coupling to curvature leads generically to nonperturbative
effects inside compact stars: compact stars acquire a scalar charge. This phenomena has been dubbed ``spontaneous
scalarization,'' since it corresponds to an instability of general-relativistic configurations~\cite{Damour:1993hw} (see
also~\cite{Pani:2010vc}). The scalar charge opens up a new channel for energy loss, through dipolar scalar
radiation~\cite{Will:1993ns}. Thus, interesting constraints on massless scalar-tensor theories arise from pulsar timing,
see e.g.~\cite{Antoniadis:2012vy} (see also~\cite{Berti:2015itd} and references therein).  We note that if the scalar is
massive, the constraints become weaker, because massive scalar fields decay exponentially $\phi(r)\approx e^{-r\mu_{\phi}}/r$, and thus the dipolar emission may be possible, but only in the late-inspiral phase~\cite{Ramazanoglu:2016kul}; if the field is too massive
the scalar is never excited.

We find that when a vector field is nonminimally coupled to gravity, as in HN gravity, compact star solutions with a
nontrivial, asymptotically vanishing vector field configuration may arise. Our results suggest that such ``vectorized''
solutions belong to two classes. One is spherically symmetric and ``induced'' by nontrivial initial conditions or
environments. We build fully nonlinear spacetimes describing such stars.  The second family may arise as the end-state
of the instability of GR solutions, and are thus ``spontaneously vectorized'' stars.

In HN gravity, the vector field is coupled to the gravitational sector, not with the matter composing the star. This
feature is of course an approximation within the interpretation (i), in which the vector field is the electromagnetic
field, while it is consistent with the interpretation (ii) of a hidden vector field, decoupled from the other matter
fields. This representation is analog to the so-called ``Jordan frame'' of scalar-tensor theory. However, while in
scalar-tensor theories the Jordan frame representation is equivalent to an ``Einstein frame'' representation, in
which the scalar field is minimally coupled to gravity and coupled to matter, there is no reason to believe that a
similar correspondence exists in vector-tensor theories, such as HN gravity, and that the theory studied in this article
admits an Einstein frame representation.  Recently, a theory with a massive scalar field minimally coupled to
gravity and nonminimally coupled to matter (i.e., an Einstein frame vector-tensor theory) has been studied
in~\cite{Ramazanoglu:2017xbl}. For the same reason discussed above, we think that there is no fundamental reason to
believe that a Jordan frame representation of such theory exists. See, however, Ref.~\cite{Ramazanoglu:2019tyi} for a thorough discussion on this issue.

\section{Hellings-Nordtvedt gravity\label{Hellings}}
In the HN gravity theory~\cite{Hellings:1973zz},
a single massless vector field is nonminimally coupled to the gravitational field.
The action for the HN theory is (henceforth we use geometric units $G=c=1$),
\begin{align}
S=&  \int d^4x\frac{\sqrt{-g}}{16\pi}(R - F_{\mu\nu}F^{\mu\nu}- {\Omega} X_{\mu} X^\mu{} R \nonumber\\
&- \eta X^{\mu} X^{\nu} R_{\mu\nu})+ S_{\rm M}\,,\label{eq:action}
\end{align}
where $X_{\mu}$ is a massless vector field, $F_{\mu\nu}=X_{\nu;\mu}-X_{\mu;\nu}$, $R_{\mu\nu}$ and $R$ are the Ricci
tensor and scalar, respectively, $\Omega$ and $\eta$\,\footnote{We choose the signs convention for the coupling constants
different from those used in Ref.~\cite{Hellings:1973zz}. Our conventions are consistent with those used in studies of scalar-tensor
theories.} are dimensionless coupling constants, and $S_{\rm M}$ is the matter fields action. The
action~\eqref{eq:action} yields the field equations \cite{Will:1993ns}
\begin{align}
\label{eq:einstein_equation}
R_{\mu\nu}-\frac{1}{2}g_{\mu\nu}R-\Omega \Theta^{(\Omega)}_{\mu\nu}-
\eta\Theta^{(\eta)}_{\mu\nu}+\Theta^{(F)}_{\mu\nu}&=8\pi G T_{\mu\nu}\,,\\
\label{eq:vector_field}
F^{\mu\nu}_{\,;\nu}+\frac{1}{2}\Omega X^{\mu}R+\frac{1}{2}\eta X^{\nu}R^{\mu}_{\nu}&=0\,,
\end{align}
where
\begin{align}
\Theta_{\mu\nu}^{(\Omega)}&=X_\mu X_\nu R +X_\alpha X^\alpha R_{\mu\nu}-\frac{1}{2}g_{\mu\nu}X_\alpha X^\alpha R\nonumber\\
&-(X_\alpha X^\alpha)_{;\mu\nu}+g_{\mu\nu}\square(X_\alpha X^\alpha)\,,\\
\Theta_{\mu\nu}^{(\eta)}&=2X^\alpha X_{(\mu} R_{\nu)\alpha}-\frac{1}{2}g_{\mu\nu}
X^\alpha X^\beta R_{\alpha\beta}\nonumber\\
&-(X^\alpha X_{(\mu})_{;\nu)\alpha}
+\frac{1}{2}\square(X_\mu X_\nu)+\frac{1}{2}g_{\mu\nu}(X^\alpha X^\beta)_{;\alpha\beta}\,\\
\Theta^{(F)}_{\mu\nu}&=-2(F^{\alpha}_{\,\,\mu}F_{\nu\alpha}-\frac{1}{4}g_{\mu\nu}F_{\alpha\beta}F^{\alpha\beta})\,,
\end{align}
and $T_{\mu\nu}$ is the matter stress-energy tensor.

We consider perturbations of static, spherically symmetric stars in HN gravity, composed by a perfect fluid. The
background is thus described by a spacetime metric with the form
\begin{equation}\label{eq:line_element}
ds^2=-Fdt^2+\frac{1}{G}dr^2+r^2d\theta^2 +r^2\sin^2\theta d\phi^2\,,
\end{equation}
where $F(r)$ and $G(r)$ are general functions of the radial coordinate $r$, and
by a stress-energy tensor with the form
\begin{equation}\label{stress_energy_tensor}
T_{\mu\nu}=(p+\rho)u_{\mu}u_{\nu}+g_{\mu\nu}p\,,
\end{equation}
where
\begin{equation}
u^{\mu}=\left(F^{-1/2},0,0,0\right)
\end{equation}
is the four-velocity of the fluid, $p(r)$ is its pressure, and $\rho(r)$ is its energy density.

We study two different equations of state (EOS) for the fluid composing the star. The first is a constant
density (CD) EOS (see, e.g.~\cite{Shapiro:1983du}) with radius $R$ and mass $M$, where $\rho=3M/(4\pi R^3)=const.$, and
\begin{align}\label{eq:axial_flat_star}
G&=1-\frac{8}{3} \pi  r^2 \rho\,,\nonumber\\
p&= \left(\frac{\rho \left(G^{1/2}-\sqrt{1-\frac{2 M}{R}}\right)}{3 \sqrt{1-\frac{2 M}{R}}-G^{1/2}}\right)\,,\nonumber\\
F&= \left(\frac{3}{2} \sqrt{1-\frac{2 M}{R}}-\frac{1}{2} G^{1/2}\right)^2 \,.
\end{align}
The second is the polytropic (Poly) EOS which has been used in~\cite{Damour:1993hw} to study
spontaneous scalarization in scalar-tensor theories,
\begin{equation}\label{eq:EOS}
\rho(p)=\left(\frac{p}{K n_0 m_b}\right)^{\frac{1}{\Gamma}} n_0 m_b+ \frac{p}{\Gamma-1}\,.
\end{equation}
where $n_0=0.1 \rm fm^{-3}=10^{53} \rm km^{-3}$, $m_b=1.66\times 10^{-24} \rm g=1.23 \times 10^{-57} \rm km$ is the average baryon mass, $\Gamma=2.34$ and $K=0.0195$ are dimensionless parameters.

\section{Linearized fluctuations of stars}\label{Lin_pert}
When $X_\mu=0$ and the field equations~\eqref{eq:einstein_equation}, \eqref{eq:vector_field} reduce to those of GR.
Therefore, all vacuum or matter solutions of GR are also solutions of the HN theory, including those describing
spherically symmetric compact stars in GR. We shall now study the stability of these solutions, considering a small
vector field perturbation:
\begin{equation}
X_\mu= \varepsilon \xi_\mu\,,
\label{eq:Xeps}
\end{equation}
where $\varepsilon\ll 1$ is a dimensionless bookkeeping parameter. At first order in $\varepsilon$,
Eq.~\eqref{eq:einstein_equation} reduces to Einstein's equations, and the vector field equation~\eqref{eq:vector_field}
can be written as
\begin{equation}
F^{\mu\nu}_{\,;\nu}-4\pi G\Omega X^{\mu}T+4\pi G\eta X^{\nu}\left(T^\mu_\nu-\frac{1}{2}\delta^\mu_\nu T\right) =0\,. \label{eq:vector_field_linear}
\end{equation}
The vector perturbation $\xi_\mu$ can be expanded in vector spherical harmonics,
\begin{equation}
\xi_{\mu}=\sum_{l} 
\begin{pmatrix}
 \begin{bmatrix}0 \\ 0 \\ a_{l}(r)(\sin\theta)^{-1}\partial_\phi Y_{l} \\
  -a_{l}(r)\sin\theta\partial_\theta Y_{l}  \end{bmatrix}& +\begin{bmatrix}
  f_{l}(r) Y_{l} \\  h_{l}(r) Y_{l} \\  k_{l}(r) \partial_\theta Y_{l} \\ k_{l}(r) \partial_\phi Y_{l} \end{bmatrix}
\end{pmatrix}e^{-\text{\rm{i}} \omega t}\,,\label{eq:vector_perturbation_def}
\end{equation}
where $Y_{lm}(\theta,\phi)$ are scalar spherical harmonics, and, since the perturbation equations do not depend on the
azimuthal index $m$, we leave that index implicit. The perturbations $a_l(r)$ (with
$l\ge1$) have axial parity, i.e. they transform as $(-1)^{l+1}$ for a parity transformation
$\theta\rightarrow\pi-\theta$, $\phi\rightarrow\phi+2\pi$, while the perturbations $f_l(r)$ and $h_l(r)$ with $l\ge0$,
and $k_l(r)$ with $l\ge1$, have polar parity, since they transform as $(-1)^l$ for a parity transformation. These two
classes of perturbations can be studied separately, because they are decoupled in the perturbations equations.

We note that when $T_{\mu\nu}=0$,
Eq.~\eqref{eq:vector_field_linear} reduces to Maxwell's equations, while the equations for the gravitational
field~\eqref{eq:einstein_equation} coincide with Einstein's equations plus terms quadratic in the vector field.
Therefore, at first order in the perturbations HN gravity coincides with Einstein-Maxwell theory for black hole (BH) spacetimes. Thus, since BHs are stable in Einstein-Maxwell theory, they are also stable against linear perturbations in
HN gravity. 

\subsection{Instabilities and spontaneous vectorization in the axial sector}
The harmonic decomposition of the linearized vector field (Eq.~\eqref{eq:vector_field_linear}) yields a system of
ordinary differential equations for the perturbation functions.
For the axial part we get (for $l\ge1$),
\begin{align}
&F G a_{l}''+\frac{1}{2}  \left(G F'+F G'\right)a_{l}'+\left[\omega^2-F\left(\frac{l (l+1)}{r^2}\right)\right]a_{l}\nonumber\\
&-2 \pi  F \left[\eta  \left( \rho - p \right)+2 \Omega  \left(\rho -3 p\right)\right]a_{l}=0\,,\label{eq:axial} 
\end{align}
where a prime denotes a derivative with respect to the coordinate $r$. The term 
\begin{equation}
\eta  \left( \rho - p \right)+2 \Omega  \left(\rho -3 p\right)\,,\label{effective_mass}
\end{equation}
in Eq.~\eqref{eq:axial} behaves as an effective mass (squared) for the vector field. When it is negative, one expects GR
configurations to be unstable against radial perturbations. For instance, in theories with $\eta=0$, this is the case
when the coupling constant $\Omega$ is negative and $\rho>3p$, or when $\Omega$ is positive and $\rho<3p$. A similar
approach has been used for a qualitative study of the stability properties of scalar-tensor theories
in~\cite{Lima:2010na,Pani:2010vc}.
\begin{figure}
\includegraphics[width=8.4cm,height=8cm,keepaspectratio]{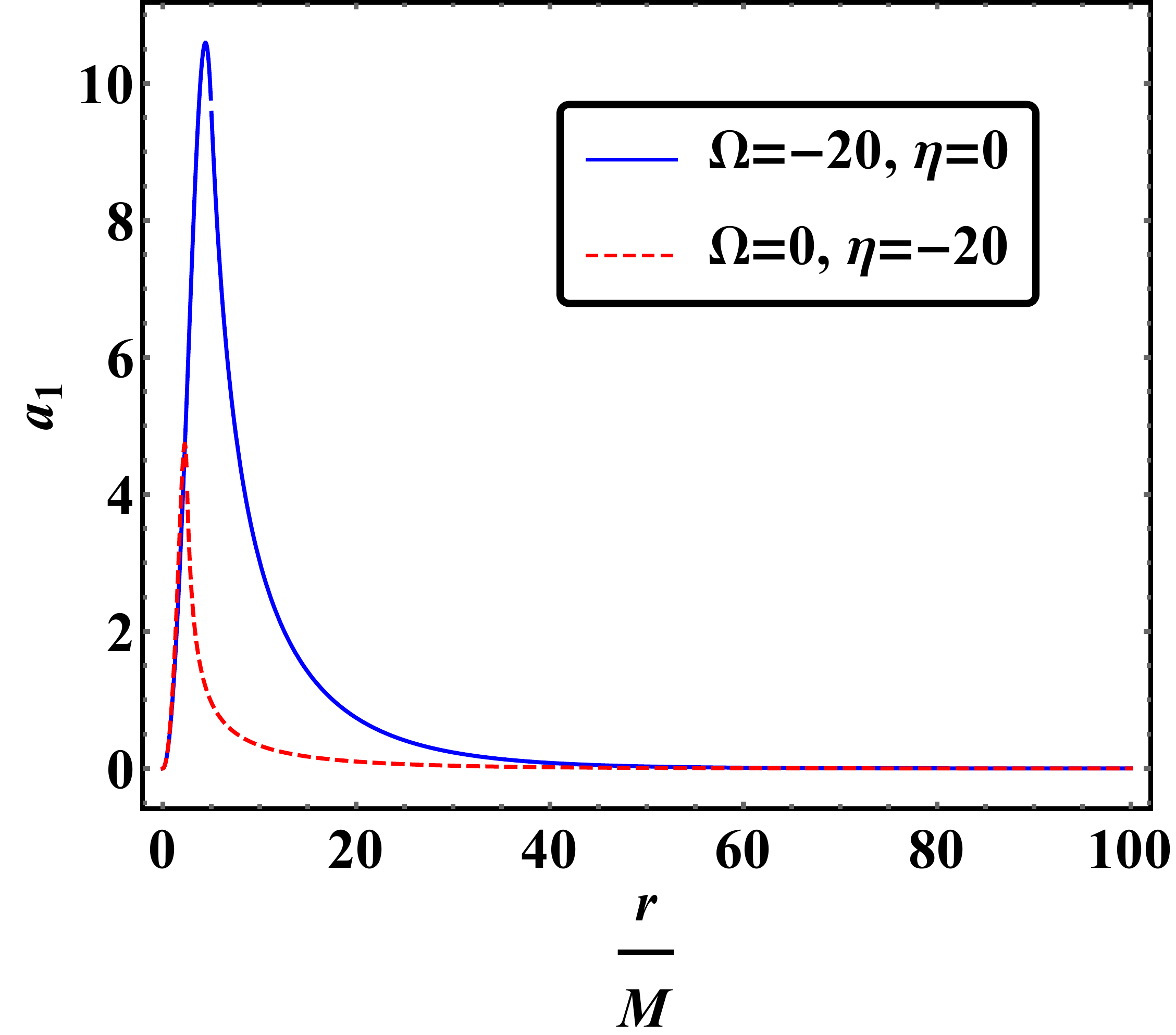}
\caption{Unstable dipolar vector perturbation profile for a CD star. The solid blue line
  ($\eta=0,\Omega=-20$) corresponds to an instability rate $M\omega=0.094 \rm i$ for a CD star compactness $M/R=0.2$. The dashed red line
  ($\Omega=0,\eta=-20$) corresponds to a rate $M\,\omega=0.071\,{\rm i}$ for a CD star compactness $M/R=0.4$.}\label{gr:plot_alm_example}
\end{figure}

We have solved numerically Eq.~\eqref{eq:axial}, for CD and Poly stars, as an eigenvalue problem
for the frequencies $\omega$.  In both cases, we have used direct integration to search for
instabilities~\cite{Macedo:2016wgh,GRIT}, looking for unstable solutions with purely imaginary frequency $\omega_{I}>0$,
and imposing regularity at the center of the star and at infinity.  Further details on the integration method are given
in Appendix~\ref{app:Axial perturbation integration}. 

\begin{figure*}
\begin{tabular}{cc}
\includegraphics[width=0.45\textwidth]{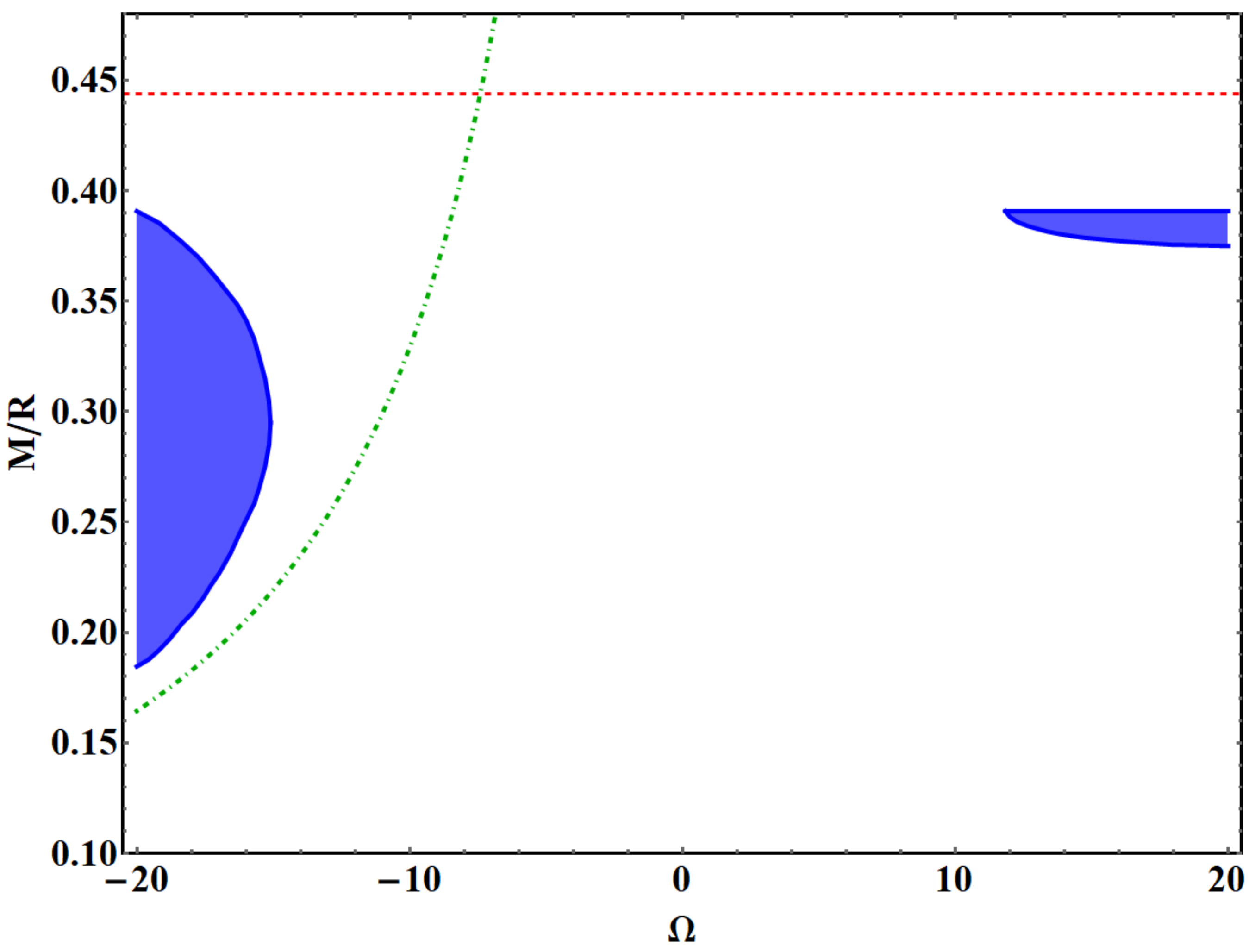}&
\includegraphics[width=0.45\textwidth]{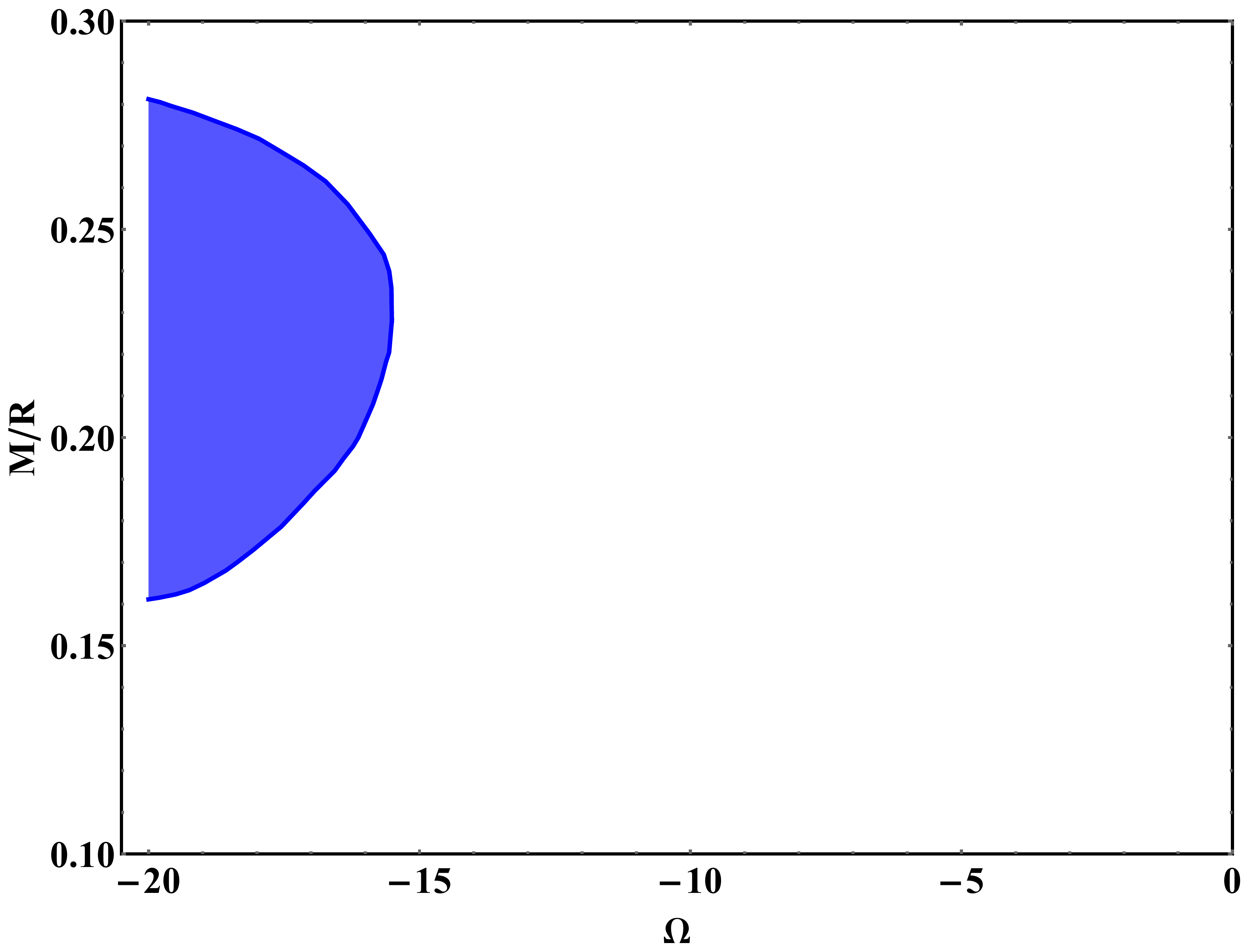}
\end{tabular}
\caption{The shaded region represents CD (left) and Poly (right) star configurations with different values
of $\Omega$ and $M/R$, which are unstable under axial perturbations in HN gravity with $\eta=0$. The
green dot-dashed curve is the Newtonian solution for a CD star, Eq.~\eqref{eq:newtonian_axial}. The dashed red line
corresponds to the Buchdal limit on the compactness of a CD star ($M/R<4/9\approx0.444$).
}
\label{gr:plot_phase_space}
\end{figure*}

We found that unstable modes are present for some configurations, the properties of which are summarized in
Figs.~\ref{gr:plot_alm_example} and \ref{gr:plot_phase_space}. Figure~\ref{gr:plot_alm_example} shows the radial profile of
dipolar ($l=1$) unstable modes for a CD star with compactness $M/R=0.2$ and couplings constants $(\Omega,\eta)=(-20,0)$
and for a CD star with compactness $M/R=0.4$ and coupling constants $(\Omega,\eta)=(0,-20)$.

When $\Omega, \eta >0$, we find unstable solutions as well. For each choice of $\eta$ and $\Omega$ we find a sequence of
characteristic frequencies corresponding to unstable solutions with nodes.

In Fig.~\ref{gr:plot_phase_space} we show the stability diagram for different values of the coupling constant $\Omega$,
assuming $\eta=0$, obtained by considering dipolar axial perturbations of stellar configurations with different values
of the compactness $M/R$.  The left panel refers to CD stars, while the right panel refers to Poly stars. The shaded region
corresponds to configurations which are unstable under axial perturbations. Strictly speaking, these regions correspond
to instability to dipolar ($l=1$) perturbations, but we find strong evidence that the configurations unstable to $l>1$
axial perturbations are also unstable to dipolar ones.  The dotted horizontal line represents the Buchdal limit
$\frac{M}{R}<\frac{4}{9}$, which we verified to be satisfied in HN theory, while the dot-dashed curve corresponds to the
Newtonian configurations for CD stars (see discussion below).  Note that, as discussed above, for negative couplings
$\Omega, \,\eta$, even Newtonian stars can become unstable. We find that unstable configurations also exists in the case
of $\eta\neq0$.

The separation between the stable and unstable regions, i.e. the boundaries of the shaded regions in
Fig.~-\ref{gr:plot_phase_space}, correspond to zero-mode solutions, i.e., static regular solutions with nonvanishing
vector field. In order to improve our understanding of this boundary, we shall now consider zero-mode solutions in the
Newtonian limit (i.e., $\frac{M}{R}\ll 1$) for a CD star. In this limit Eq.~\eqref{eq:axial} reduces to
\begin{equation}
a_{l}''-\left(\frac{l(l+1)}{r^2}+\mu^2\right) a_{l}=0\,,\label{eq:axnewt}
\end{equation}
where $\mu^2=2 \pi \rho(2\Omega+\eta)$ is the effective mass of the vector inside the star. Imposing regularity at the
origin, the general solution of Eq.~\eqref{eq:axnewt} inside the star is (modulo an arbitrary multiplicative constant)
$a_l=\sqrt{r}J_{l+1/2}(-i\mu r)$, with $J_\nu$ Bessel function.  Outside the star $\rho=0$ and imposing regularity at
infinity Eq.~\eqref{eq:axnewt} gives $a_l\propto r^{-l}$. Matching the interior and the exterior solution at the radius
of the star $r=R$ we find that a regular solution exists only for $I_{l-1/2}(-i\mu R)=0$, where $I_\nu$ is the modified
Bessel function, i.e., for $\mu R=i\pi$, which corresponds to
\begin{equation}\label{eq:newtonian_axial}
-3 M \left(\eta +2 \Omega \right)= 2 \pi ^2 R\,.
\end{equation}
When $\eta=0$, this equation admits a nontrivial solution for negative values of $\Omega$. Thus, for $\eta=0$,
$\Omega<0$ we expect the presence of unstable solutions.

The Newtonian prediction~\eqref{eq:newtonian_axial} is shown in Fig.~\ref{gr:plot_phase_space} (green dot-dashed
curve). We note that this line is close to the boundary of the unstable region, and it is closer for smaller values of
the compactness, as expected.  At fixed negative coupling constant $\Omega$, as the compactness increases to large
values, the quantity $\rho-3P$ decreases. For Poly stars, at $M/R\sim 0.27$ it becomes negative. The effective
squared mass of the vector \eqref{effective_mass} is then positive, and the star is stable. Thus, all the main features
of Fig.~\ref{gr:plot_phase_space} can be understood in simple terms. For the same reasons, unstable solutions lying on
the right side of the plot (positive coupling constants) exist for very large values of the compactness, when the
effective mass squared is again negative.

It is worth noting that our results resemble those of scalar-tensor theories (compare our Fig.~\ref{gr:plot_phase_space}
with Fig. 1 of Ref.~\cite{Pani:2010vc}). The root of the mechanism is the same: a tachyonic instability that is either
triggered by a ``wrong'' sign of the coupling constants or by the wrong sign of the trace of the stress-energy tensor.
Despite these similarities, we are here discussing the dipolar axial sector excitations of the vector field, which have
a very different behavior from those of the scalar field. The end state of this instability is unknown to us.

Including backreaction on Einstein's equations, the axial perturbations give a contribution to
the $(\theta,\theta)$ component of Einstein's equations, which can be considered as an effective stress-energy tensor.
This suggests that  the star will be made to rotate as a result of such instability. Another outcome is possible: that
the star exits the instability window through mass shedding. The fate of stars on the unstable branch remains an open
issue.

\subsection{Spontaneous and induced vectorization in the polar sector}
%
\begin{figure}
\includegraphics[width=8.4cm,height=8cm,keepaspectratio]{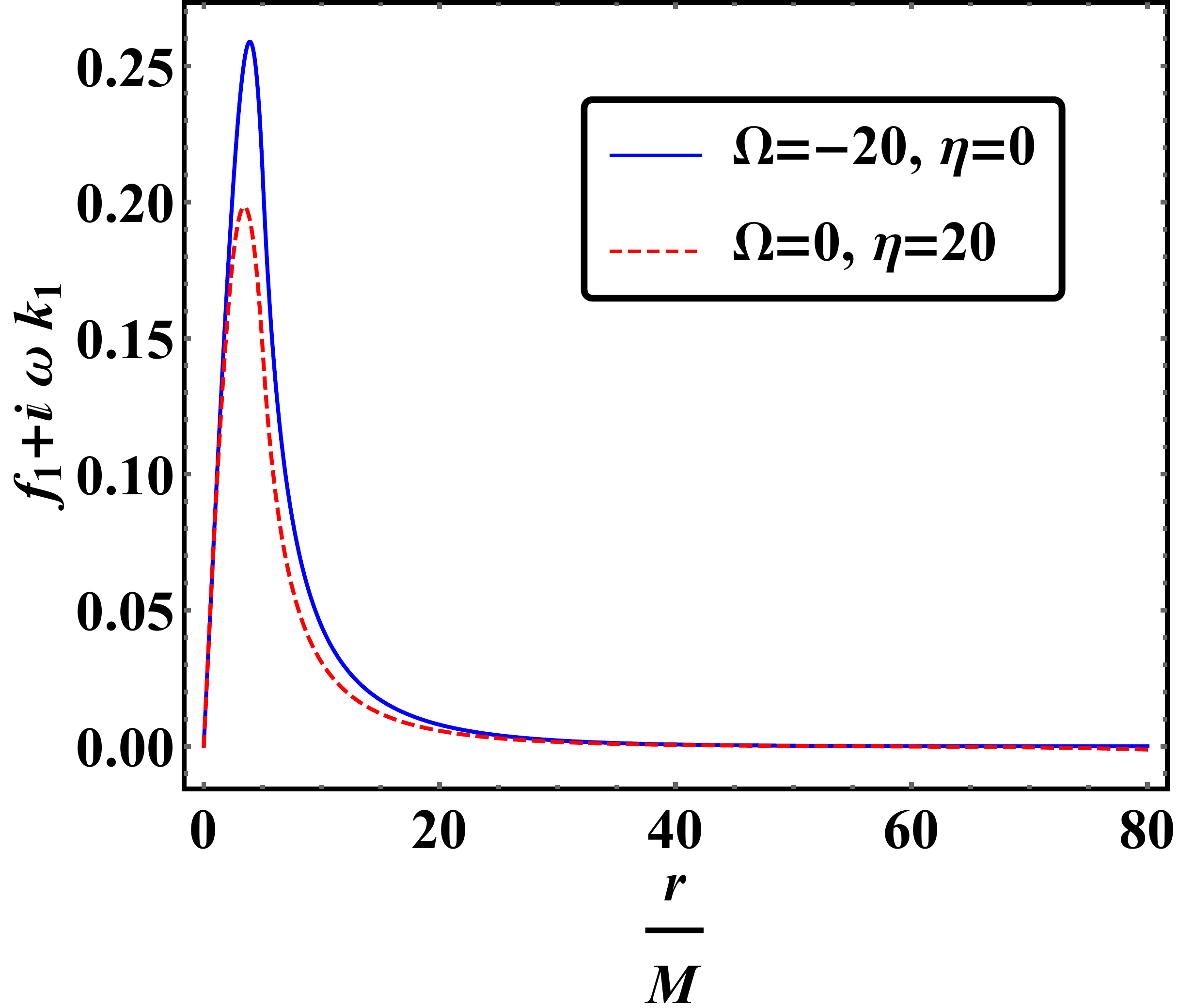}
\caption{Unstable dipolar vector perturbation profile for a constant density star with compactness $M/R=0.2$. The solid blue line ($\eta=0,\Omega=-20$) corresponds to an instability rate $M\omega=0.103 \rm i$. The dashed red line ($\Omega=0,\eta=20$) corresponds to a rate $M\,\omega=0.0989 \rm i$.}\label{gr:plot_psi_example}
\end{figure}
Since we are interested in static and spherically symmetric solutions of the full non linear field equations in HN
gravity, we shall now study linear vector field perturbations with polar parity in this theory.

\noindent {\bf Dynamical case.} To begin with, let us consider monopolar ($l=0$) perturbations. 
In the exterior of the star, we find that the $l=0$ polar perturbation equations reduce to those of GR, i.e.
\begin{align}
  &(\mathrm{i}\omega h_0+f_0')\left(rFG'+4FG-rGF'\right)\nonumber\\
  &+2rGF\left(\mathrm{i}\omega h_0+f_0'\right)'=0\,,\nonumber\\
&\omega\left(\mathrm{i}\omega h_0+f_0'\right)=0\,.\label{eq:l0gr}
\end{align}
Since the radial electric field $E_r$ is proportional to $\mathrm{i} \omega h_0+f_0'$, when $\omega\neq0$ the second of
Eqs.~\eqref{eq:l0gr} implies that $E_r=0$, and thus the wave is pure gauge: there are no spherically symmetric electromagnetic waves
with radial electric fields in the exterior of the star, in HN gravity as in GR. Then, since the solution inside the
star has to match the exterior solution, $E_r$ is pure gauge in the entire spacetime. In other words, there is no
dynamical {\it linear} instability for spherically symmetric modes.

Let us now consider the polar perturbations with the $l\ge1$ case. By replacing the
expansion~\eqref{eq:vector_perturbation_def} in the $r$ component of Eq.~\eqref{eq:vector_field_linear} we find:
\begin{equation}
h_l=\frac{-l(l+1) F k_l'+\mathrm{i} r^2 \omega f_l'}{F \left(2 \pi  r^2
   ((6 \Omega +\eta ) p-(\eta +2 \Omega ) \rho )-l(l+1)\right)+r^2 \omega^2}.\label{eq:l1r}
\end{equation}
Replacing Eq.~\eqref{eq:l1r} in the $t$ and $\theta$ components of Eq.~\eqref{eq:vector_field_linear} we obtain a system
of coupled ordinary differential equations (ODEs) in $f_l$ and $k_l$, that is shown in Appendix \ref{app:dipolar}, Eqs.~\eqref{eq:dipolar_one} and \eqref{eq:dipolar_two}. 

\begin{figure}
\includegraphics[width=8.4cm,height=8cm,keepaspectratio]{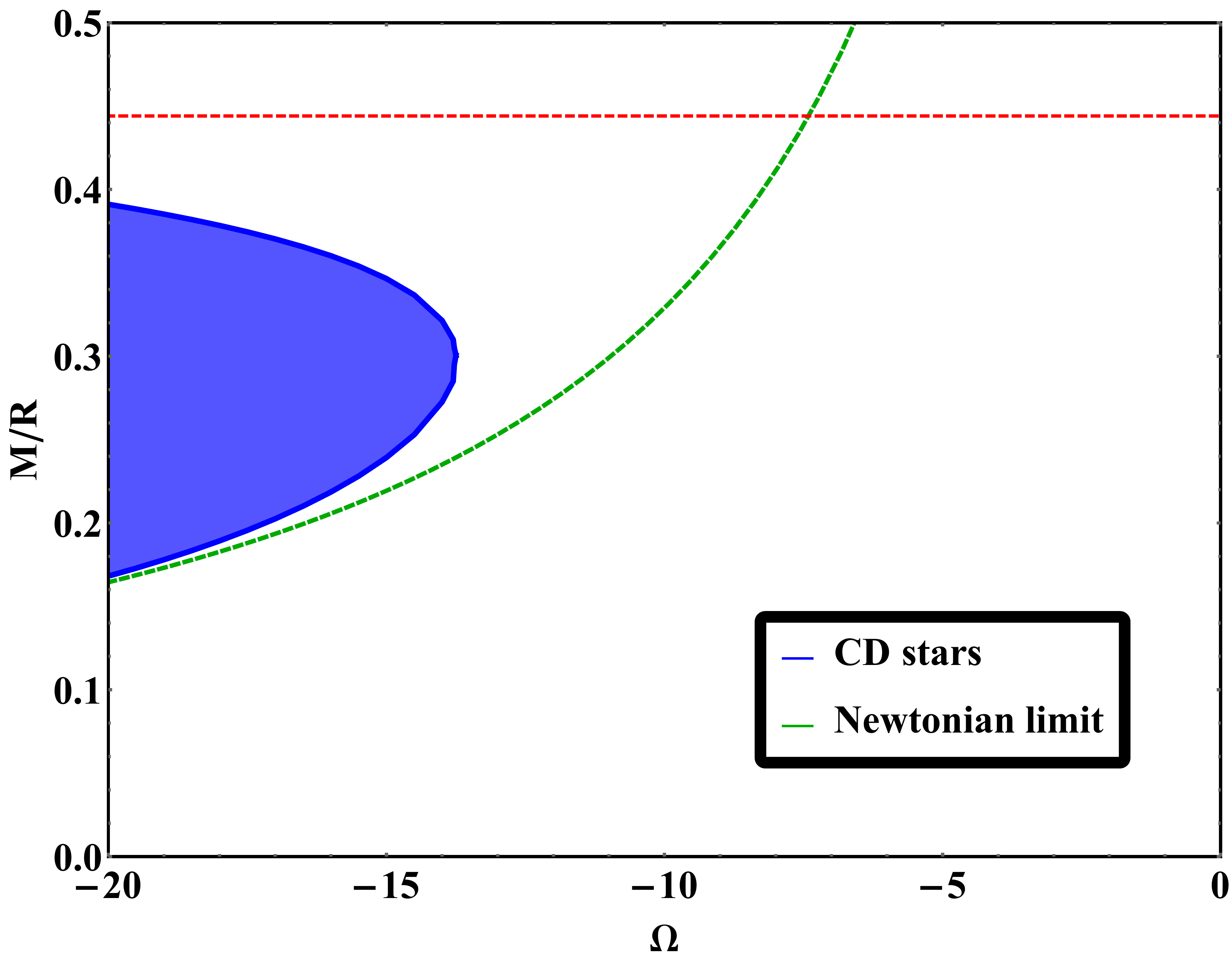}
\caption{Instability diagram for CD stars, for polar perturbations with $l=1$. The shaded region represents solutions
  which are unstable under polar perturbations in HN gravity with $\eta=0$. The green dot-dashed curve describes
  zero-frequency modes in the Newtonian regime, Eq.~\eqref{eq:newtonian_polar_lone}. The dashed red line corresponds to
  the Buchdal limit ($M/R<4/9\approx0.444$).}\label{gr:plot_phase_space_polar_CDS}
\end{figure}

The perturbation equations have simpler expressions in the exterior of the star. Indeed, as $\rho=p=0$ 
Eqs.~\eqref{eq:dipolar_one} and \eqref{eq:dipolar_two} can be cast as a single ``master equation'' in terms of the quantity
\begin{equation}
\psi = f_l + \mathrm{i} \omega k_l\,,
\end{equation}
which is:
\begin{align}\label{eq:schwarz_polar}
& \frac{\left(l^2+l\right) \psi  }{2 M r-r^2}+\frac{l (l+1) (2 M-r) \psi '' }{l (l+1)
   (2 M-r)+r^3 w^2}\nonumber\\
   &+\frac{2 l (l+1)  \left(l (l+1) (r-2 M)^2-M r^3
   w^2\right) \psi '}{r \left(l (l+1) (2 M-r)+r^3 w^2\right)^2}=0\,.
\end{align}
Solving the Cauchy problem given by Eqs.~\eqref{eq:dipolar_one} and \eqref{eq:dipolar_two}, with appropriate initial
conditions (as discussed in Appendix \ref{app:Axial perturbation integration}) and matching at the boundary of the star
with Eq.~\eqref{eq:schwarz_polar}, we find unstable configurations for CD stars. In Fig.~\ref{gr:plot_psi_example},
for instance, we show the radial profile of an unstable mode with $l=1$ for a constant density star of compactness
$M/R=0.2$, for $\Omega=-20, \eta=0$ or $\eta=20, \Omega=0$.
As in the case of axial perturbations, we can construct the instability diagram in the space ($\Omega, M/R$). The static
zero-mode solutions in the Newtonian limit $M/R\ll1$ yields
\begin{equation}\label{eq:newtonian_polar_lone}
3 M \left(\eta -2 \Omega \right)= 2 \pi ^2 R\,.
\end{equation}
In Fig.~\ref{gr:plot_phase_space_polar_CDS} we show the instability region (shaded region) in the ($\Omega, M/R$) plan,
for $\eta=0$ and negative values of the coupling constant $\Omega$. The horizontal dotted line represents the Buchdal
limit, and the dot-dashed curve represents the Newtonian zero-mode solutions corresponding to
Eq.~\eqref{eq:newtonian_polar_lone}.

An interesting feature of these results is that the contribution of the coupling $\eta$ to the effective mass squared
has opposite sign from that of axial perturbations: large negative $\eta$ make Newtonian stars unstable against axial
perturbations, and large positive $\eta$ turn Newtonian stars unstable against polar perturbations.

\begin{center}
\begin{figure}\includegraphics[width=8.4cm,height=8cm,keepaspectratio]{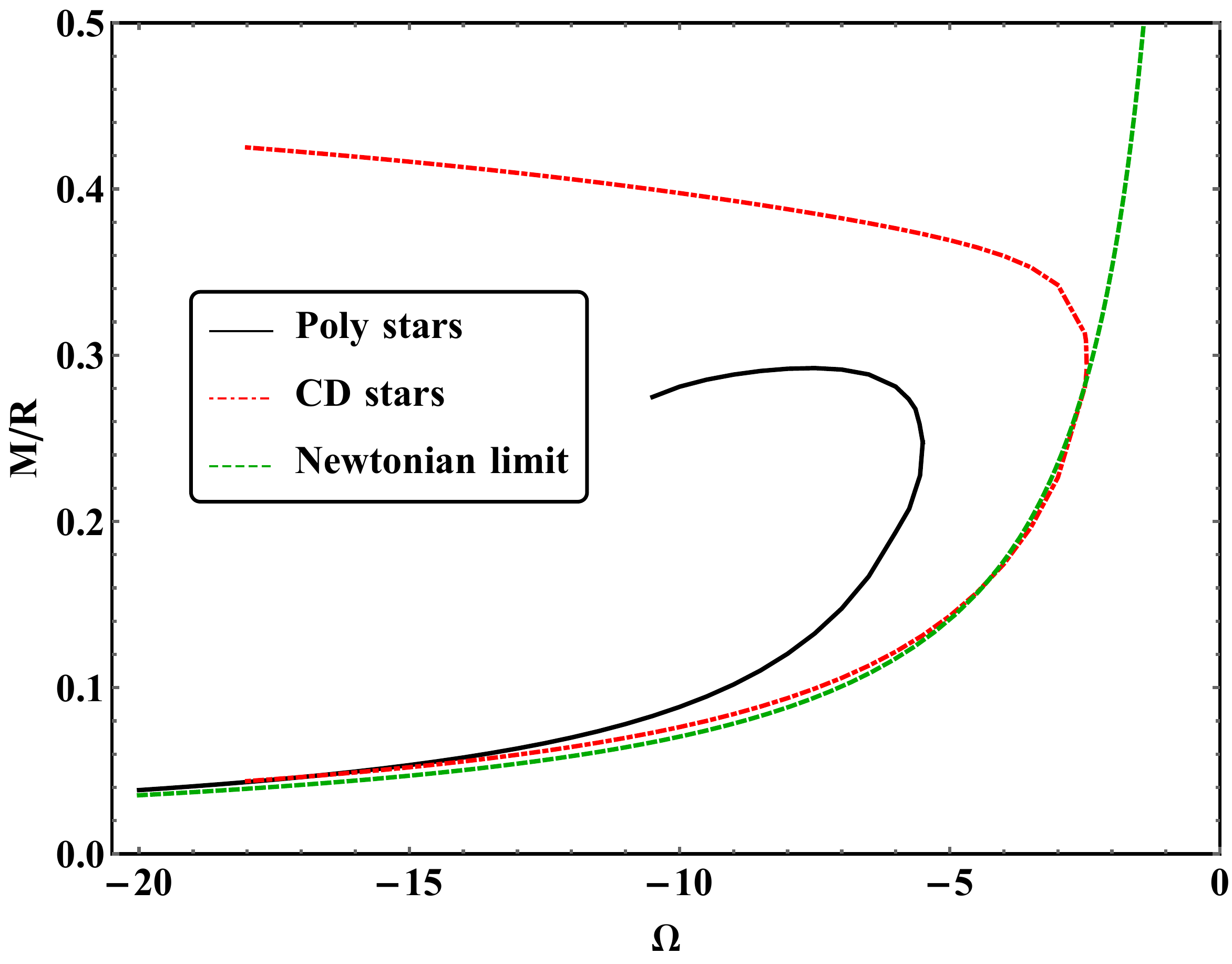}
\caption{Linear static vector field solutions for a NS background (solid black line) and for a CD star (dot-dashed red line)
 for $\eta=0$. The dashed green line corresponds to the Newtonian analytic solution for the polar sector in the CD star
 background.}
\label{gr:plot_linearsol_CDS_EOS}
\end{figure}
\end{center}

\noindent {\bf Static case.} We showed that spherically symmetric polar modes have no interesting dynamics. 
However, there is still room for the existence of nontrivial static ($\omega=0$) solutions. Replacing the
expansion~\eqref{eq:vector_perturbation_def} in the field equations~\eqref{eq:vector_field_linear} we find: 
\begin{equation}
h_0=0\,,
\end{equation}
and\footnote{It is possible to show that $k_0$ can be canceled out by the use of an appropriate combination of the independent  components of the modified Maxwell equation.},
\begin{align}
\big[&-f_0 '  \left(2 r \left(m' -2\right)+r (r-2 m ) \nu ' +6 m \right)+2 r (r-2 m ) f_0 '' \nonumber\\
&+4 \pi  r^2 (2 \Omega  (3 p-\rho )+\eta  (3 p+\rho ))f_0\big]\frac{e^{-\nu  }}{2 r^2}=0\,.\label{eq:static_linear_polar_equation}
\end{align}
Solving Eq.\eqref{eq:static_linear_polar_equation} we find a class of linear static vector field solutions, for both the
CD and the Poly star configurations. Equation~\eqref{eq:static_linear_polar_equation} also implies an analytical relation
between the compactness and the coupling constant, in the Newtonian regime for a CD star. Indeed, since $h_0=0$, we can
assume the following form for the vector field:
\begin{equation}
X_\mu=(f_0(r)/r,0,0,0)\,.\label{eq:one_component_assumption}
\end{equation}
In the limit $M/R\ll1$, for a CD star Eq.~\eqref{eq:static_linear_polar_equation} reduces to
\begin{equation}
f_{0}''-\mu^2 f_{0}=0\,,
\end{equation}
where $\mu^2=2 \pi \rho(2\Omega-\eta)$ is the effective mass of the vector inside the star. Imposing regularity at the
origin we find $f_0=e^{\mu r}-e^{-\mu r}$ inside the star, and imposing regularity at infinity we find $f_0=const.$ in
the exterior. By matching the interior and exterior solutions at the radius of the star we find
\begin{equation}
6M(\eta -2 \Omega)=\pi ^2 R\,.
\end{equation}
The static solutions are summarized in Fig.\ref{gr:plot_linearsol_CDS_EOS}. These solutions might be said to be {\it induced},
rather than arising spontaneously as the end product of an instability: they arise as the end product of (perhaps special) initial conditions.
Such vectorized solutions have no parallel in scalar-tensor theory and do not exist in the axial sector of HN gravity
itself. Note again that $\eta$ contributes with the opposite sign, relative to axial perturbations in Eq.~\eqref{eq:axnewt}.
From these solutions we can conclude that in GR, a vector field coupled with the curvature of the spacetime can have a
nontrivial profile around compact stars. Moreover, this result suggests that vectorized stars can appear even in
full nonlinear HN gravity. 

We should mention that, generically, the vector $X_\mu$ will tend to grow {\it all} its components. Thus, with the exception
of a measure-zero set of initial conditions, finding only a nonzero timecomponent is impossible. In other words, the spherically symmetric state that occurs at linear (and nonlinear, as we show below) level is not generic and should always be accompanied by the linear instability of the nonsymmetric modes. However, from a purely mathematical level the distinction between induced and spontaneous processes can be made and we have adopted such nomenclature here.

\section{Static, vectorized neutron stars in the HN theory}

\subsection{Formalism}
We shall now determine full nonlinear, stationary and spherically symmetric NS configurations in HN theory, solutions
of Eqs.~\eqref{eq:einstein_equation} and \eqref{eq:vector_field}. In other words, we show that the induced vectorized
solutions, found above at a linear level, do indeed exist at full nonlinear level.  Hereafter, we assume $\eta=0$. For
convenience, we rewrite the line element of Eq.~\eqref{eq:line_element} defining $F=e^{\nu(r)}$ and
$G=1-\frac{2m(r)}{r}$. A spherically symmetric vector field can only have nonvanishing $t$- and $r$-
components. Moreover, the $r$ component of the vector field equation reduces to
\begin{align}
\label{eq:X1}
&\frac{X_r}{4 r^3}\Big[\Omega (r-2 m)  (-2 m' (r \nu '+4)+(4 r-6 m) \nu ')\nonumber\\
&+2 r (r-2 m) \nu ''+r (r-2 m) \nu '^2\Big]=0
\end{align}
which implies $X_r\equiv0$. 
Therefore, all the space components of the vector field identically vanish:
\begin{equation}
X_{\mu}=\left\{X(r),0,0,0\right\}\,.
\end{equation}
The structure equations for the star are given by the $(t,t)$, $(r,r)$ components of the Einstein equations
\eqref{eq:einstein_equation}, the vector field equation \eqref{eq:vector_field} and the conservation of the
stress-energy tensor:
\begin{equation}\label{eq:stress_energy}
\nabla_{\nu}T^{\mu\nu}=0\,.
\end{equation}
We note that Eq.~\eqref{eq:stress_energy} holds in HN gravity because the GR modifications do not affect the matter
section of the action~\eqref{eq:action}; therefore, as explicitly shown in Ref.~\cite{Will:1993ns}, the four-divergence
of the stress-energy tensor vanishes in this theory, as in GR. We thus obtain a system of four ODEs in the variables
\begin{equation}
\left\{m(r),\nu(r),p(r),X(r)\right\}\,.
\end{equation}
These modified Tolman-Oppenheimer-Volkoff (TOV) equations are shown explicitly in Appendix~\ref{app:TOV}, Eq.~\eqref{eq:TOV_system}; their expansion near
the center of the star is shown in Appendix~\ref{app:exp_center}.

We numerically solve the modified TOV equations, assuming the polytropic EOS introduced in Eq.~\eqref{eq:EOS}. At the
surface of the star (where the pressure vanishes) we evaluate the components of the spacetime metric, of the vector
field and of its first derivative. Then, we numerically integrate the equations in the exterior, which correspond to
the modified TOV equations with $\rho=p=0$, from the stellar surface to infinity.
With this procedure, we have a unique solution of the modified TOV equations for any choice of the quantities
\begin{equation}
\left\{p_{c},X_{c},\Omega\right\}\,,
\end{equation}
i.e. for any choice of the pressure and of the (time component of the) vector field at the center of the star, and for
any value of the coupling constant $\Omega$. At infinity, the vector field has the form
\begin{equation}
X_0(r\gg R)=X_{\rm \infty}+\frac{\alpha}{r}\,,\label{eq:field_at_inf}
\end{equation} 
where $\alpha$ is a constant which can be considered as a sort of vector charge (although it is not a Noether
charge, as in the case of the scalar charge in scalar-tensor theories~\cite{EspositoFarese:2004cc}), and $X_{\rm\infty}$
is the asymptotic value of (the time component of) the vector field.

We search for solutions of the modified TOV equations with a nontrivial vector field configuration, and with the same
asymptotic behavior as GR solutions, i.e., $X_{\rm\infty}=0$.
Is it worth noting that the mass function does not remain constant in the exterior of the star, due to the energy
contribution of the nontrivial vector field.  The gravitational mass that a far away observer can measure, i.e., the Arnowitt-Deser-Misner
mass of the spacetime, is the asymptotic value of $M(r)$. This is the definition of gravitational mass that we are going
to use in the rest of the paper.

The baryonic mass of the NS is defined as~\cite{Misner:1974qy}
\begin{equation}
\bar{m}=m_b \int d^3x \sqrt{-g}u^0 n(r)\,,
\end{equation}
where $u^0$ is the time component of the four-velocity and $n(r)$ is the number density of baryons, which is related to
the pressure by
\begin{equation}
p(r)=K n_0 m_b \left(\frac{n(r)}{n_0}\right)^{\Gamma}\,.
\end{equation}
For each vectorized solution, we can evaluate the normalized binding energy of the stars defined as,
\begin{equation}
\frac{E_b}{M}=\frac{\bar{m}}{M}-1\,.\label{eq:Ebm}
\end{equation}
In order to have a bound object, we need $E_b$ to be positive. Moreover, the dependence of the gravitational mass on the
central density often conveys information on the stability of the configuration. Indeed, in GR a necessary condition for
radial stability of a stellar configuration is $dM/d\rho_c<0$, or equivalently
$dM/dR>0$~\cite{Friedman:1988er,Shapiro:1983du}. The condition for stability in generalized theories depends on the
number of extra fields, and becomes more complicated~\cite{Brito:2015yfh}.
\subsection{Vectorized stars}\label{Vectorized stars}
%
\begin{figure}
\begin{center}\includegraphics[width=8.5cm,height=8cm,keepaspectratio]{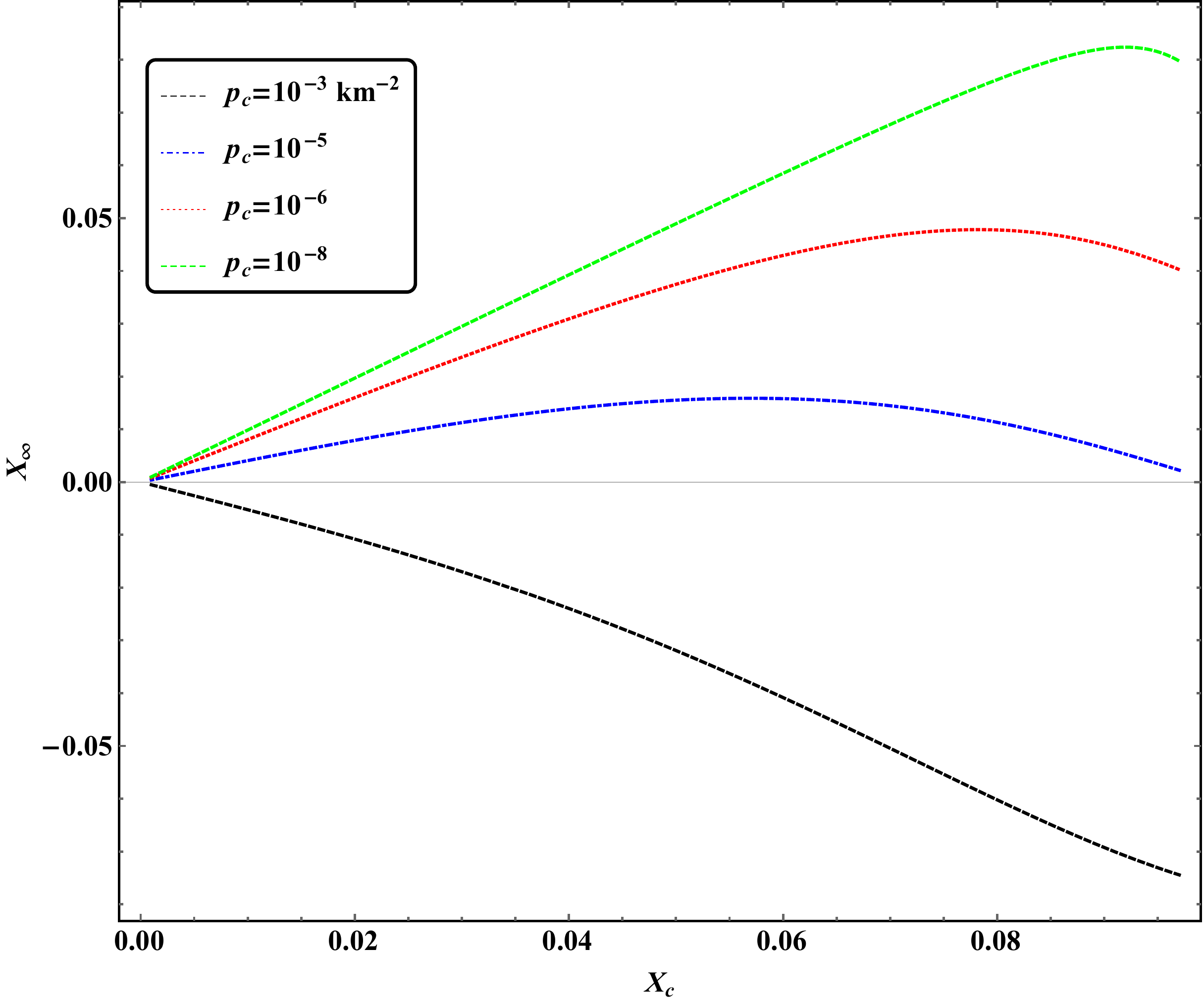}
\caption{Time component of the vector field at infinity, as a function of its value at the center of the star, for
  $\Omega=-5$ and for different values of the central pressure. From top to bottom,
  $p_c=10^{-8},10^{-7},10^{-6},10^{-5},10^{-4},10^{-3}\,{\rm Km}^{-2}$.}\label{gr:plot_field_inf_diff_pressure}
\end{center}
\end{figure}
In Fig.~\ref{gr:plot_field_inf_diff_pressure} we show the results of the numerical integration of
the modified TOV equations in the case of $\Omega=-5$. In the figure the asymptotic value of the vector field,
$X_{\rm\infty}$, is shown as a function of the vector field at the center, $X_c$. Each curve corresponds to a different
value of the central pressure $p_c$. The ``physical'' solutions, i.e., those with the same asymptotic behavior as the
GR solutions, correspond to the intersections of the curves with the $X_{\rm\infty}=0$ axis. We see that all curves
intersect the $X_{\rm\infty}=0$ axis at the origin (corresponding to the GR solution), but some of them also have a
second intersection, which corresponds to the vectorized solutions.

%
\begin{figure}
\begin{center}
\includegraphics[width=8.4cm,height=7cm,keepaspectratio]{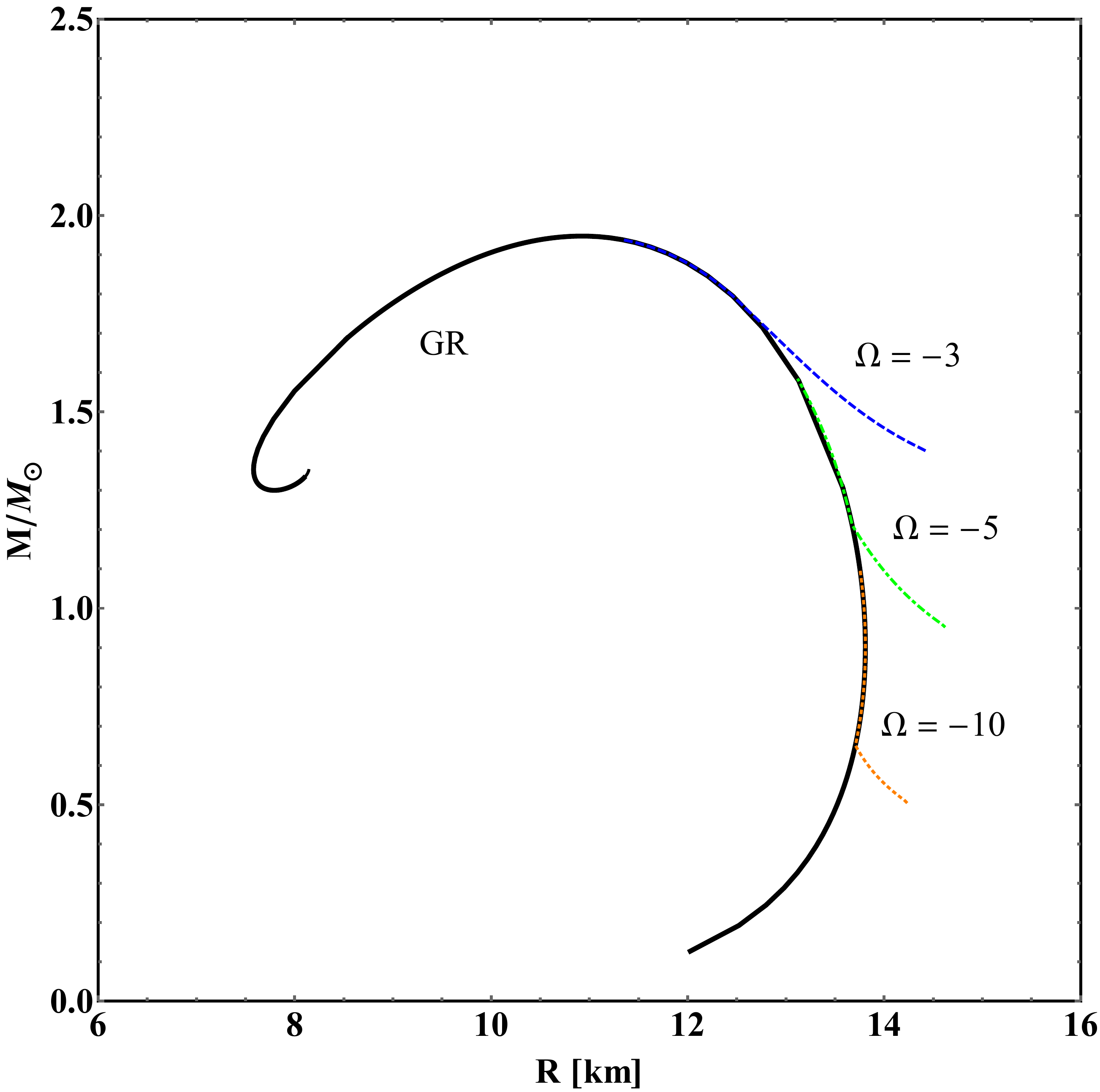}
\caption{Mass-radius configurations for different coupling constants. The longest line (black) represents the solution
  for NSs in GR given the EOS in Eq.~\eqref{eq:EOS}, while the other branches correspond to vectorized
  solutions.}\label{gr:mass_radius}
\end{center}
\end{figure}
\begin{figure}
\begin{center}
\includegraphics[width=8.4cm,height=8cm,keepaspectratio]{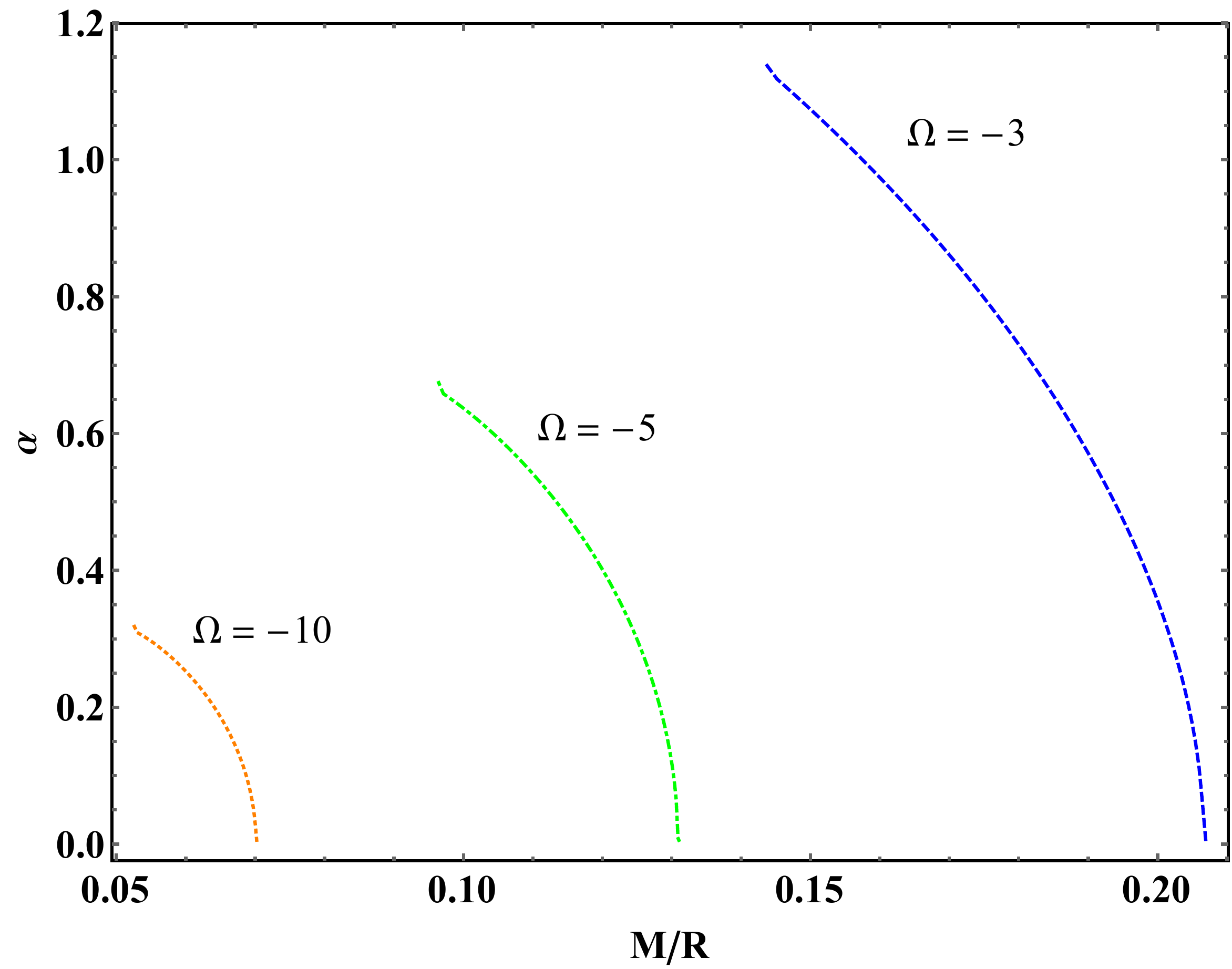}
\caption{Vector field charge as a function of the compactness, for different stellar configurations. All the stars carrying a zero charge ($\alpha=0$) are always a solution of the theory, although they are not explicitly shown in the figure. In fact, in the range in which there are stars with a nonzero charge, the solutions are always two, as it was clear from Fig.\ref{gr:plot_field_inf_diff_pressure}.}\label{gr:plot_vectorizedsol_chargevscompactness}
\end{center}
\end{figure}
We computed the vectorized solutions for a wide range of values of the central pressure and of the
coupling constant $\Omega$.  The masses and the radii of these configurations are shown in Fig.~\ref{gr:mass_radius};
the corresponding values of the vector charge $\alpha$ (see Eq.~\eqref{eq:field_at_inf}) is shown in
Fig.~\ref{gr:plot_vectorizedsol_chargevscompactness} as a function of the compactness. 
We can see that comparing different compact star configurations (for a given value of the coupling $\Omega$), as the
compactness {\it decreases} there is a smooth transition between GR stars and vectorized solutions. Then, below a
threshold value of the compactness, the vectorized solution does not exist anymore. As discussed in Appendix
\ref{app:exp_center}, this behavior is due to the fact that, when the compactness reaches the threshold value, the
modified TOV equation are not well behaved near the center of the star, since they become degenerate. Below the
threshold value, the weak energy condition is violated, leading to an unphysical object.
%
%
For positive values of $\Omega$, our linearized results (see Sec.~\ref{Lin_pert}) suggest that there is more than one
solution corresponding to every value of the central pressure, with different numbers of nodes in the profile of $X_0(r)$.


%
\begin{figure}
\begin{center}
\includegraphics[width=8.4cm,height=8cm,keepaspectratio]{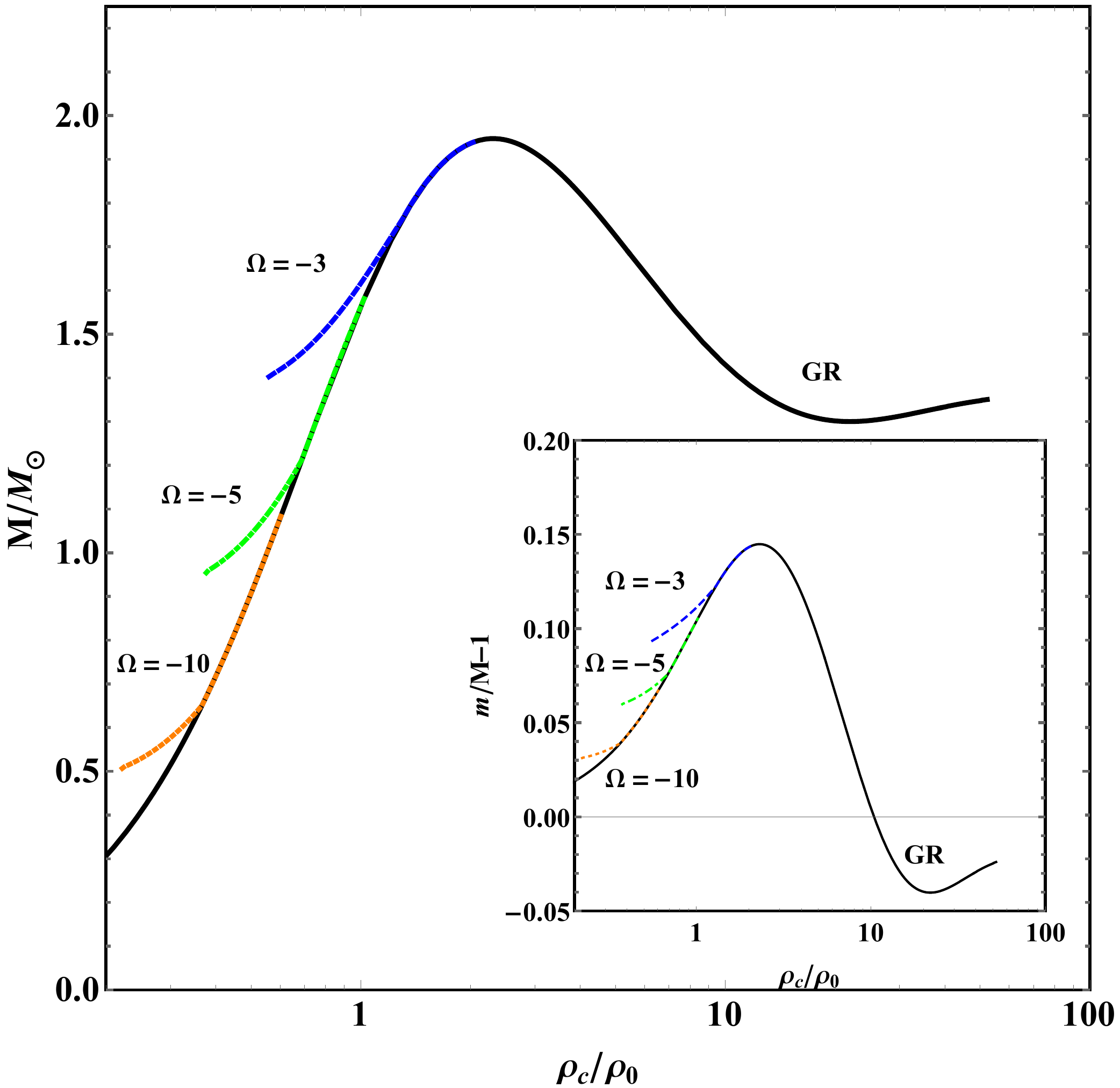}
\caption{Gravitational mass as a function of the (normalized) central baryonic density. This shows again that the solutions for $\Omega<0$ are stable solutions of the theory, since they are associated with $d M/d\rho_c<0$. (Inset) The normalized binding energy as a function of the central energy density.}\label{gr:plot_massenergy}
\end{center}
\end{figure}
We did not perform a dynamical stability analysis of such vectorized solutions. However, important information is conveyed by the dependence of the total (normalized) gravitational mass on the central energy density ($\rho_c=m_b n(0)$). This function is shown in  Fig.~\ref{gr:plot_massenergy}.
In GR, a criterium for stability is that $d M/d\rho_c<0$. Such criterium holds only approximately in modified
theories~\cite{Harada:1998ge,Brito:2015yfh}. We will use this as merely indicative, as more sophisticated analysis tools include a dynamical evolution or the analysis done in Ref.~\cite{Brito:2015yfh}. With such criterium, all the vectorized
solutions associated with a negative coupling constant are stable: they are in the stable branch of the $dM/d\rho_c$
curve. Moreover, in the inset of Fig.~\ref{gr:plot_massenergy} we can see how the solutions in vector-tensor theory are associated with larger binding energy than in GR, indicating that they are the preferred
configuration. For positive couplings, however, the behavior is the opposite.

Finally, we note that the instability of solutions in the positive $\Omega$ semiplane is consistent with previous results in scalar-tensor gravity~\cite{Mendes:2016fby}. Thus, none of the solutions associated with a positive coupling constant are stable and most likely do not play any astrophysical role.

Somewhat surprisingly, our results show that the changes in the NS structure with respect to the GR are smaller for {\it
  larger values of the coupling constants}. This finding is consistent with previous reported results in a related
theory~\cite{Ramazanoglu:2017xbl}. In fact, it is apparent from Fig.~\ref{gr:plot_vectorizedsol_chargevscompactness}
that the charge (and so the field) inside vectorized stars is larger for smaller (absolute) values of the
compactness. This is probably due to the role of the coupling constant $\Omega$ in the modified TOV system. In fact, it resembles
the role of a mass for the scalar field in scalar-tensor gravity: the larger the scalar mass, the smaller are the
modifications from GR.  Taking into account this, from the mass-radius plot one can quantify the range of values of
$\Omega$ that allow vectorized stars to exist,
\begin{equation}
\Omega\approx [-12,-2]\,.
\end{equation}

Finally, let us stress that there is no {\it linear} instability of spherically symmetric stars in GR.
Thus, there is no linear mechanism to drive a GR star to these new vectorized states that we just described.
Such vectorized spherical stars must therefore arise out of nonlinear effects (such as selected initial conditions).

\section{Discussion}
We have shown that very simple extensions of the Einstein-Maxwell theory allow for nontrivial phenomena to exist.  Once
curvature couplings are allowed, general relativistic stars are allowed but generically unstable. We are unable at this
point to follow the evolution of such instability or even to understand its end state, since it drives a nonspherically
symmetric mode.  The end state could be a star with a nontrivial vector field, but it could also simply be a GR solution
away from the instability region.

We find interesting novel spherically symmetric star configurations in this theory. They do not seem to arise out of any
``spontaneous-vectorization'' mechanism but are rather induced by initial conditions. These stars carry a nonzero electric charge and give rise to dipolar electromagnetic radiation when accelerated. The calculation of such fluxes and its use in astrophysical
observations to constrain the coupling constants is an interesting open problem.

\acknowledgments
We are indebted to Clifford Will for useful correspondence on some aspects of the Hellings-Nordtvedt theory.
V.C. acknowledges financial support provided under the European Union's H2020 ERC 
Consolidator Grant ``Matter and strong-field gravity: New frontiers in Einstein's 
theory'' Grant Agreement No. MaGRaTh--646597.
This project has received funding from the European Union's Horizon 2020 research and innovation programme under the Marie Sklodowska-Curie Grant Agreement No. 690904.
We acknowledge financial support provided by FCT/Portugal through Grant No. PTDC/MAT-APL/30043/2017.
We acknowledge the SDSC Comet and TACC Stampede2 clusters through NSF-XSEDE Grant No. PHY-090003.
The authors would like to acknowledge networking support by the GWverse COST Action 
CA16104, ``Black holes, gravitational waves and fundamental physics.'' L.A. acknowledges financial support provided by Funda\c{c}ao para a Ci\^{e}ncia e a Tecnologia Grant No. PD/BD/128232/2016 awarded in the framework of the Doctoral Programme IDPASC-Portugal.


\appendix

\section{Equations for polar perturbations}\label{app:dipolar}
The perturbation equations with polar parity are
\begin{widetext}
\begin{align}\label{eq:dipolar_one}
&\frac{G f_l''  (1-\frac{r^2 \omega^2}{F (2 \pi  r^2
   ((6 \Omega +\eta ) p -(\eta +2 \Omega ) \epsilon  )-l (l+1))+r^2
   \omega^2})}{F}+\frac{i l (l+1) \omega G
   k_l'' }{F (l (l+1)+2 \pi  r^2 ((-\eta -6 \Omega ) p -(-\eta
   -2 \Omega ) \epsilon  ))-r^2 \omega^2}\nonumber\\
&-\frac{f_l  (l (l+1)+2 \pi  r^2 (3 (-\eta -2 \Omega ) p +(-\eta +2 \Omega )
   \epsilon  ))}{r^2 F}-\frac{i l (l+1) \omega k_l }{r^2
   F}\nonumber\\
&+\frac{i l (l+1) \omega k_l' }{2 r F (F (2 \pi 
   r^2 ((6 \Omega +\eta ) p +(-\eta -2 \Omega ) \epsilon  )-l (l+1))+r^2   \omega^2)^2} (r F (-G F' 
   (l (l+1)\nonumber\\
   &+2 \pi  r^2 ((-\eta -6 \Omega ) p -(-\eta -2 \Omega ) \epsilon
    ))-r^2 \omega^2 G' )+r^3 (-\omega^2) G
   F' +F^2 (r G'  (l (l+1)+2 \pi  r^2 ((-\eta -6
   \Omega ) p \nonumber\\
   &-(-\eta -2 \Omega ) \epsilon  ))+4 G (l (l+1)+\pi
    r^3 ((6 \Omega +\eta ) p' +(-\eta -2 \Omega ) \epsilon
   ' ))))\nonumber\\   
&-\frac{f_l' }{2 r F (F (2 \pi  r^2 ((6 \Omega
   +\eta ) p +(-\eta -2 \Omega ) \epsilon  )-l (l+1))+r^2 \omega^2)^2} (r F (G (l^2 (l+1)^2
   F' \nonumber\\
   &+4 \pi  r^2 ((-\eta -6 \Omega ) p  (F'  (l
   (l+1)-2 \pi  r^2 (-\eta -2 \Omega ) \epsilon  )+2 r \omega^2)-(-\eta -2 \Omega
   ) \epsilon   (l (l+1) F' +2 r \omega^2)\nonumber\\
   &+\pi  r^2 (-\eta -6 \Omega
   )^2 p ^2 F' +\pi  r^2 (-\eta -2 \Omega )^2 \epsilon  ^2
   F' +r^2 \omega^2 ((-\eta -6 \Omega ) p' -(-\eta -2 \Omega ) \epsilon
   ' )))+r^2 \omega^2 G'  (l (l+1)\nonumber\\
   &+2 \pi  r^2 ((-\eta -6
   \Omega ) p -(-\eta -2 \Omega ) \epsilon  )))+r^3 \omega^2 G
   F'  (l (l+1)\nonumber\\
   &+2 \pi  r^2 ((-\eta -6 \Omega ) p -(-\eta -2 \Omega )
   \epsilon  ))-F^2 (r G' +4 G)
   (l (l+1)+2 \pi  r^2 ((-\eta -6 \Omega ) p -(-\eta -2 \Omega ) \epsilon
    ))^2)=0\,,\\
    \label{eq:dipolar_two}
&+\frac{i \omega
   G  f_l'' }{F  (2 \pi  r^2 ((6 \Omega +\eta )
   p -(\eta +2 \Omega ) \epsilon  )-l (l+1))+r^2 \omega^2}+\frac{i \omega f_l }{r^2 F }\nonumber\\
   & -\frac{G  k_l''  (2 \pi  F  ((6 \Omega +\eta )
   p +(-\eta -2 \Omega ) \epsilon  )+\omega^2)}{F  (2 \pi  r^2 ((6
    \Omega +\eta ) p +(-\eta -2 \Omega ) \epsilon  )-l (l+1))+r^2 \omega^2}+\frac{k_l
    (2 \pi  F  ((-\eta -6 \Omega ) p -(-\eta -2 \Omega )
   \epsilon  )-\omega^2)}{r^2 F }\nonumber\\
   &+\frac{i \omega f_l'  }{2 r F 
   (F  (2 \pi  r^2 ((6 \Omega +\eta ) p +(-\eta -2 \Omega )
   \epsilon  )-l (l+1))+r^2 \omega^2)^2}(r F  (G  F'  (l
   (l+1)\nonumber\\
   &+2 \pi  r^2 ((-\eta -6 \Omega ) p -(-\eta -2 \Omega ) \epsilon  ))+r^2
   \omega^2 G' )+r^3 \omega^2 G  F' +F ^2 (4
   G  (\pi  r^3 ((-\eta -6 \Omega ) p' \nonumber\\
   &-(-\eta -2 \Omega ) \epsilon
   ' )-l (l+1))-r G'  (l (l+1)+2 \pi  r^2 ((-\eta -6 \Omega
   ) p -(-\eta -2 \Omega ) \epsilon  ))))\nonumber\\
    & -\frac{k_l'  }{2 r F  (F  (2 \pi  r^2 ((6
   \Omega +\eta ) p +(-\eta -2 \Omega ) \epsilon  )-l (l+1))+r^2 \omega^2)^2} (r \omega^2 F  (G  F'  (l
   (l+1)\nonumber\\
   &+4 \pi  r^2 ((6 \Omega +\eta ) p +(-\eta -2 \Omega ) \epsilon  ))+r^2
   \omega^2 G' )-F ^2 (-2 \pi  r G  F' 
   ((-\eta -6 \Omega ) p \nonumber\\
   &-(-\eta -2 \Omega ) \epsilon  ) (l (l+1)+2 \pi  r^2
   ((-\eta -6 \Omega ) p -(-\eta -2 \Omega ) \epsilon  ))+r \omega^2 G' 
   (l (l+1)+4 \pi  r^2 ((-\eta -6 \Omega ) p \nonumber\\
   &-(-\eta -2 \Omega ) \epsilon
    ))+4 l (l+1) \omega^2 G )+r^3 \omega^4 G  F' +2
   \pi  F ^3 (r G'  ((-\eta -6 \Omega ) p -(-\eta -2 \Omega )
   \epsilon  ) (l (l+1)\nonumber\\
   &+2 \pi  r^2 ((-\eta -6 \Omega ) p -(-\eta -2 \Omega )
   \epsilon  ))+2 l (l+1) G  (r (-\eta -6 \Omega ) p' +2 (-\eta
   -6 \Omega ) p \nonumber\\
   &-r (-\eta -2 \Omega ) \epsilon ' -2 (-\eta -2 \Omega ) \epsilon   )))=0\,.
\end{align}
\end{widetext}
\section{Numerical integration of the perturbation equations}\label{app:Axial perturbation integration}
We here discuss the numerical integration of perturbation
equations of static, spherically symmetric stars in HN gravity.
In order to enforce a regular behavior near the center of the star, we perform an asymptotic expansion of the
axial perturbation equation \eqref{eq:axial}, by expanding the perturbation function as
%
\begin{equation}
a_{l}=\sum_{i=0}^{N}a_l^i r^i\,.
\end{equation}
We truncate the expansion at $N=4$ because we found that further coefficient does not affect significantly the results.
Thus, we find the values of the coefficients $a_l^{i>0}$ in terms of $a_l^0$ (which can be set to an arbitrary
value). In terms of these coefficients we can compute $a_l(r_0)$ and $a_{l,r}(r_0)$ at $r_0\ll R$. We then numerically
integrate Eq.~\eqref{eq:axial} from $r_0$ to the surface of the star $r=R$ and, imposing regularity of the
perturbations, from the surface to $r\gg R$.

Unstable modes have frequency $\omega=\omega_R+ \rm{i} \omega_I$, with $\omega_{I}>0$. Since we look for the onset of
the instability, we look for solutions with purely imaginary frequency, i.e., $\omega_R=0$, by matching the solution far
away from the star with
%
%
%
%
\begin{equation}
a_{l}(r)\approx e^{\omega_{\text{\rm{I}}}  r}c_1+e^{-\omega_{\text{\rm{I}}} r}c_2\,,
\end{equation}
where $c_1$ and $c_2$ are two constants of integration. Finally, since we require an asymptotically flat spacetime, we
impose $c_1=0$. We thus find a perturbation which grows in time and regular at spatial infinity, behaving asymptotically as
%
\begin{equation}
a_{l}(t,r)=c_2e^{-\omega_{\text{I}} r} e^{\omega_{\text{I}} t}\,.
\end{equation}
In order to solve the perturbation equations with polar parity~\eqref{eq:dipolar_one} and \eqref{eq:dipolar_two}, we follow
the same approach. The only difference is that, for each value of the harmonic index $l$, we have two perturbation
functions, $f_l(r)$ and $k_l(r)$.
\section{Modified TOV equations}\label{app:TOV}
In the following, we show the full nonlinear system of equations that describes static, nonrotating stars in HN gravity:
\begin{widetext}
\begin{align}\label{eq:TOV_system}
  &\frac{e^{-\nu}}{2 r^2} \big[-4 e^{2 \nu} (4 \pi  r^2 \rho-m')-4 r^2 X'^2+-e^{\nu }
    \Omega  X^2 (4 m' (r \nu '+1)+4 (3 m-2 r) \nu '+r (r-2 m) (\nu '^2-4 \nu ''))\nonumber\\
    &-e^{\nu } (4 \Omega  X (r (r-2 m) X''-X' (r m'+2 r (r-2 m) \nu '+3 m-2r))+2 r (2
    \Omega -1) (r-2 m) X'^2)\big]=0\,,\nonumber\\
  &\frac{e^{-\nu }}{2 r^2 (r-2 m)} \big[2 m (-2 r^2 (4 e^{\nu }-1) X'^2-2 e^{\nu }
    (r \nu '+1)-2 r \Omega  X (r \nu '+4) X'+\Omega  X^2(r \nu ' (r \nu '+2)-2))\nonumber\\
&    +16 r m^2 e^{\nu }X'^2+r^2 (-16 \pi  r p e^{\nu }+2 \Omega  X (r \nu '+4) X'+2 e^{\nu }
    (\nu '+2 r X'^2)-2 r X'^2-\Omega  X^2 \nu ' (r \nu '+2))\big]=0\,,\nonumber\\
&\frac{e^{-\nu }}{4 r^2} \big[-2 X' (2 r (m'-2)+r (r-2 m) \nu '+6 m)\nonumber\\
&+\Omega  X (-2 m' (r \nu '+4)+2 (2 r-3 m) \nu '+r (r-2 m) (2 \nu
   ''+\nu '^2))+4 r (r-2 m) X''\big]=0\,,\nonumber\\
&\frac{(r-2 m) (2 p'+(p+\rho) \nu ')}{2 r}=0\,,
\end{align}
\end{widetext}
%
This system is invariant if under the transformation
\begin{equation}\label{eq:invariance_TOV}
\{\nu(r)\rightarrow\nu_0+\tilde{\nu}(r),\, X\rightarrow e^{\frac{\nu_0}{2}}\tilde{X}\}\,,
\end{equation}
where $\nu_0$ is an arbitrary constant.
%
%
%
%
\section{Expansion at the center of the modified TOV equations}\label{app:exp_center}
The asymptotic expansion near the center of the star of the modified TOV equations~\eqref{eq:TOV_system} is:
%
\begin{align}
\rho&= \rho_c+\rho_1 r +\frac{\rho_2}{2}r^2\,,\nonumber\\
p (r)&= p_c+p_1 r +\frac{p_2}{2}r^2\,,\nonumber\\
\nu (r)&= \nu_c+\nu_1 r +\frac{\nu_2}{2}r^2\,,\label{eq:nu_expansion}\\
X (r)&= {X}_{c}+{X}_{1} r +{X}_{2}\frac{r^2}{2}\,,\nonumber\\
m (r)&= m_3 r^3\nonumber\,.
\end{align}
Performing the transformation~\eqref{eq:invariance_TOV}, we solve the modified TOV equations for $\tilde{\nu}=\nu-\nu_c$
and $\tilde{X}$. Comparing order by order, the nonvanishing coefficients of the expansion~\eqref{eq:nu_expansion} are
\begin{align}\label{eq:central_values}
  m_3 &= \frac{4 \pi  \left(3 p_c \tilde{X}_{c}^2 \Omega  (\Omega +2)+\rho_c
    \left(\tilde{X}_{c}^2 \Omega  (2 \Omega +1)-1\right)\right)}{9 \tilde{X}_{c}^2
    \Omega ^2 \left(\tilde{X}_{c}^2 (\Omega -1)+1\right)-3}\,,\nonumber\\
  \tilde{\nu}_2 &= \frac{8 \pi  \Big(3 p_c \left(\tilde{X}_{c}^2 \Omega  (2 \Omega +1)
    -1\right)}{9 \tilde{X}_{c}^2 \Omega ^2 \left(\tilde{X}_{c}^2 (\Omega -1)+1\right)-3}\nonumber\\
  &+\frac{8 \pi  \rho_c \left(\tilde{X}_{c}^2 \Omega  (4 \Omega -1)-1\right)\Big)}{9
    \tilde{X}_{c}^2 \Omega ^2 \left(\tilde{X}_{c}^2 (\Omega -1)+1\right)-3}\,\nonumber,\\
  \tilde{X}_{0,2}&=\frac{4 \pi  \tilde{X}_{c} \Omega  \left(3 \tilde{X}_{c}^2 \Omega
    (3 p_c+\rho_c)+3 p_c-\rho_c\right)}{9 \tilde{X}_{c}^2 \Omega ^2 \left(\tilde{X}_{c}^2 (\Omega -1)+1\right)-3}\,,\nonumber\,\\
  p_2&=-\frac{4 \pi  (p_c+\rho_c) \Big(3 p_c \left(\tilde{X}_{c}^2 \Omega  (2 \Omega +1)
    -1\right)}{9 \tilde{X}_{c}^2 \Omega ^2 \left(\tilde{X}_{c}^2 (\Omega -1)+1\right)-3}\nonumber\\
  &-\frac{4 \pi  (p_c+\rho_c) \rho_c \left(\tilde{X}_{c}^2 \Omega  (4 \Omega -1)-1\right)
    \Big)}{9 \tilde{X}_{c}^2 \Omega ^2 \left(\tilde{X}_{c}^2 (\Omega -1)+1\right)-3}\,.
\end{align}
In the limit $\Omega=0$ these coefficients reduce to those of the TOV equations in GR. Moreover, we note that as
\begin{equation}
9 \tilde{X}_{c}^2    \Omega ^2 \left(\tilde{X}_{c}^2 (\Omega -1)+1\right)-3=0\label{eq:den}
\end{equation}
all the coefficients 
defined in \eqref{eq:central_values}
diverge. Thus, when Eq.~\eqref{eq:den} admits solution $r=\bar r$ inside the star, i.e. $0\le\bar r\le R$, then the
modified TOV equations do not allow for a regular vectorized solution. Since, as the compactness of the star decreases, the root $\bar r$ becomes smaller, this is the reason for the existence of a threshold compactness under which the vectorized solution disappears (see e.g. Fig.~\ref{gr:plot_vectorizedsol_chargevscompactness}).

\bibliography{Bibliography}

\end{document}